\documentclass[aps, prl, superscriptaddress, reprint,nobibnotes]{revtex4-2}
\usepackage{amsmath}
\usepackage{color}
\usepackage{comment}
\usepackage{bm,graphicx,hyperref}
\hypersetup{%
  breaklinks = {true},
  citecolor = {blue},
  colorlinks = {true},
  linkcolor = {red},
}
\usepackage{lineno}
\usepackage[bottom]{footmisc}

\begin{document}

\title{Electronic Self-passivation of Single Vacancy in Black Phosphorus via a Controlled Ionization}


\author{Hanyan Fang}
\thanks{These authors contributed equally to this work.}
\affiliation{Department of Chemistry, National University of Singapore, Singapore 117543, Singapore}

\author{Aurelio Gallardo}
\thanks{These authors contributed equally to this work.}
\affiliation{Institute of Physics, Academy of Sciences of the Czech Republic, Prague, Czech Republic}

\affiliation{Department of Condensed Matter Physics, Faculty of Mathematics and Physics, Charles University, V Holešovičkách 2, Prague 180 00, Czech Republic}

\author{Dikshant Dulal}
\affiliation{Yale-NUS College, 16 College Avenue West, Singapore 138527, Singapore}

\author{Zhizhan Qiu}
\affiliation{Department of Chemistry, National University of Singapore, Singapore 117543, Singapore}

\author{Jie Su}
\affiliation{Department of Chemistry, National University of Singapore, Singapore 117543, Singapore}

\author{Mykola Telychko}
\affiliation{Department of Chemistry, National University of Singapore, Singapore 117543, Singapore}

\author{Harshitra Mahalingam}
\affiliation{Yale-NUS College, 16 College Avenue West, Singapore 138527, Singapore}

\author{Pin Lyu}
\affiliation{Department of Chemistry, National University of Singapore, Singapore 117543, Singapore}

\author{Yixuan Han}
\affiliation{Department of Chemistry, National University of Singapore, Singapore 117543, Singapore}

\author{Yi Zheng}
\affiliation{Zhejiang Province Key Laboratory of Quantum Technology and Device, Department of Physics, Zhejiang University, Hangzhou 312007, China}

\author{Yongqing Cai}
\affiliation{Joint Key Laboratory of Ministry of Education, Institute of Applied Physics and Materials Engineering, University of Macau, Taipa, Macau 999078, China}

\author{Aleksandr Rodin}
\thanks{Corresponding author}
\affiliation{Yale-NUS College, 16 College Avenue West, Singapore 138527, Singapore}

\affiliation{Centre for Advanced 2D Materials (CA2DM), National University of Singapore, Singapore 117543, Singapore}

\author{Pavel Jel\'{\i}nek}
\thanks{Corresponding author}
\affiliation{Institute of Physics, Academy of Sciences of the Czech Republic, Prague, Czech Republic}

\affiliation{Regional Centre of Advanced Technologies and Materials, Czech Advanced Technology and Research Institute (CATRIN), Palacký University Olomouc,Olomouc 78371, Czech
Republic}

\author{Jiong Lu}
\thanks{Corresponding author}
\affiliation{Department of Chemistry, National University of Singapore, Singapore 117543, Singapore}

\affiliation{Centre for Advanced 2D Materials (CA2DM), National University of Singapore, Singapore 117543, Singapore}

\date{\today}

\begin{abstract}

We report that mono-elemental black phosphorus presents a new electronic self-passivation scheme of single vacancy (SV). By means of low-temperature scanning tunneling microscopy and bond-resolved non-contact atomic force microscopy, we demonstrate that the local reconstruction and ionization of SV into negatively charged $\mathrm{SV}^-$ leads to the passivation of dangling bonds and thus the quenching of in-gap states, which can be achieved by mild thermal annealing or STM tip manipulation. SV exhibits a strong and symmetric Friedel oscillation (FO) pattern, while $\mathrm{SV}^-$ shows an asymmetric FO pattern with local perturbation amplitude reduced by one order of magnitude and a faster decay rate. The enhanced passivation by forming $\mathrm{SV}^-$ can be attributed to its weak dipole-like perturbation, consistent with density-functional theory and numerical calculations. Therefore, self-passivated $\mathrm{SV}^-$ is electronically benign and acts as a much weaker scattering center, which may hold the key to further enhance the charge mobility of BP and its analogs.    

\end{abstract}	

\maketitle

High-mobility two-dimensional semiconductors (2DSCs) are essential for the development of ultra-thin high-speed and energy-efficient electronics and optoelectronics~\citep{Li2014,Qiao2014,Buscema2014,chen2015,sheng2021}. The intrinsic mobility of defect-free 2DSCs is normally set by the effective mass of carriers and phonon scattering processes~\citep{Chen2008,Yu2014}. However, materials synthesis and device fabrication processes of 2DSCs, including metal chalcogenides and black phosphorus (BP), inevitably introduce surface vacancies with dangling bonds (DBs) due to the volatile nature of chalcogen and phosphorus (P) atoms~\citep{Lin2015,Schuler2019,Liu2017,Kiraly2017}. Such atomic defects often act as undesirable sinks for charge carriers and nonradiative recombination centers of photoexcited electron-hole pairs~\citep{Pashley1993,Hong2015,Khan2017}, which becomes one of the major device performance limiting factors. Therefore, effective passivation of vacancies in high-mobility 2DSCs is vital to maintaining their high-performance device characteristics.

The ideal surface passivation method should deactivate only the defect states of 2DSCs without a permanent crystal lattice change and degradation of their electronic performance. Inspired by the conventional passivation technologies used in the semiconductor industry, various strategies, including chemical functionalization~\citep{Yu2014,Han2016}and surface coating~\citep{Pak2015,Park2017}, have been exploited for the passivation of surface vacancies in 2DSCs to remove the associated detrimental in-gap electronic states. However, most passivation schemes developed to date mainly improve the photoluminescence quantum yield without significant enhancement in charge transport properties~\citep{Amani2015,Han2016}, and even degrade the electronic performance by altering the van der Waals structure~\citep{Cho2015,Pak2015}.

Here, we demonstrate that mono-elemental BP, a prototypical high-mobility 2DSC with a unique puckered square lattice ~\citep{Li2014,Luo2015,Li2015,Deng2017a,Li2017,Jung2020}, is able to effectively passivate isolated vacancies by a self-driven lattice reconstruction process to form negatively charged $\mathrm{SV}^-$ sites. Such an  self-passivation mechanism of vacancies and the associated in-gap electronic states  relies on the formation of homoelemental hypervalent bonding, which is not reported in heteroelemental 2DSCs (e.g., metal chalcogenides). 

\begin{figure}
    \centering
    \includegraphics[width=\columnwidth]{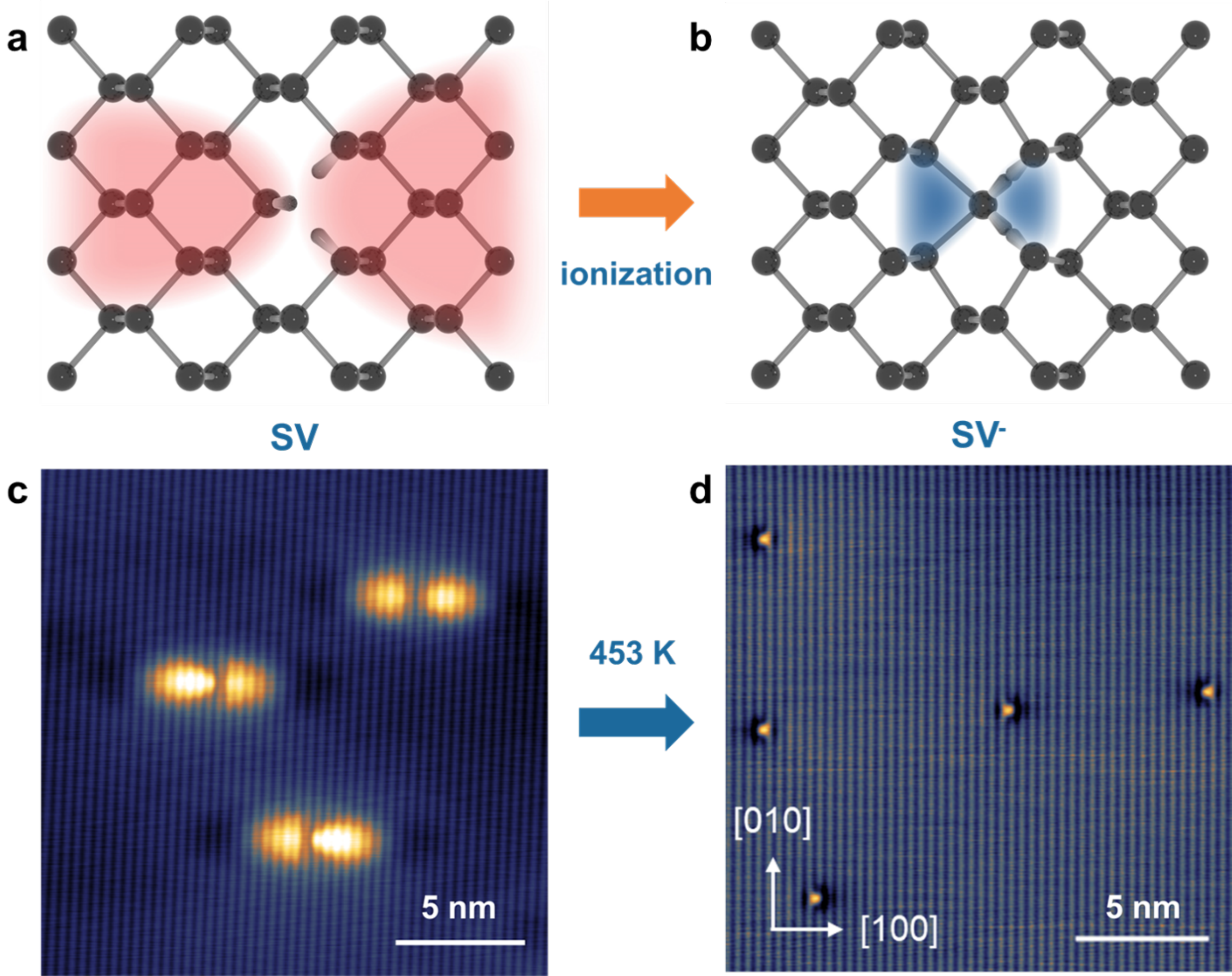}
    \caption{\textbf{Thermally driven self-passivation of SV into $\mathbf{SV^-}$ via local reconstruction and ionziation.} (a) and (b) Schematics highlight unsaturated SV with dangling bondns and self-passivated ($\mathrm{SV}^-$), respectively. (c, d) STM images of BP surface taken before and after the thermal annealing at 453 K followed by rapid cooling. The crystallographic directions are indicated in (d). STM setpoints: $V_S = -1.0$ V, $I = 0.3$ nA.}
    \label{fig:Generation}
\end{figure}

As illustrated in Fig.~\ref{fig:Generation}a, removal one P atom from the buckled lattice in BP creates an SV and leaves DBs at three adjacent P sites . Neutral SVs in BP are predicted to possess in-gap electronic states, which play an essential role in modulating electronic and optical properties of BP and thus affecting their device characteristics~\citep{Xia2014,Wang2015a,Liu2016,Kiraly2017,Riffle2018,Babar2019}. Despite advances in BP research, the atomic-scale structural and electronic properties of SVs and their impact on charge dynamics in BP remain elusive. In addition, the microscopic knowledge of defect passivation mechanism for eliminating DBs in SVs of BP is still missing.

To this end, we have exploited low-temperature scanning tunneling microscopy (STM) and bond-resolved non-contact atomic force microscopy (nc-AFM) to probe the electronic self-passivation of intrinsic SV in BP. The versatile bond configuration of P atoms and mono-elemental composition of BP introduce a new deactivation mechanism of SV via local reconstruction, saturating the DBs at adjacent P sites. The structural transformation and migration of SV in BP can proceed readily due to a relatively low energy barrier ~\citep{Cai2016,Yao_2020}. Amongst various SV configurations predicted previously ~\citep{Liu2014,Guo2015,Hu2015a,Hu2015,Gaberle2018}, this one involves the formation of hypervalent four-coordinated P atoms at the defect center (Fig.~\ref{fig:Generation}b), allowing for the saturation of all the DBs but leaving one extra electron at SV sites (denoted as negatively charged $\mathrm{SV}^-$). The lattice flexibility of BP ensures a feasible transformation of SVs into electronically inactive ones via self-passivation, which can be triggered by a mild thermal annealing or tip-induced local ionization.

 STM imaging of BP surface cleaved in-situ reveals a ubiquitous presence of defects with large-sized dumbbell-shaped appearance across more than tens of zigzag (ZZ) lattice chains (Fig.~\ref{fig:Generation}c), which are tentatively labeled as neutral SV (in the following context, SV represents neutral SV unless stated otherwise). The spatially extended dumbbell feature is probably associated with the delocalized bound hole states over SV due to its shallow acceptor nature, which will be further discussed in detail below. We also performed constant-height nc-AFM imaging of these defects with a CO-functionalized tip to monitor the spatial variation of frequency shift ($df$) in the Pauli repulsive regime. The higher $df$ reflects positions of atoms or chemical bond with a high electron density \citep{Gross1110}. The nc-AFM image of SV shows the perfect ZZ chains attributed to the P atoms at the topmost surface (Fig.~\ref{fig:SV_UnderSurface}). This result suggests that these SVs reside in the sub-surface layers, consistent with a recent DFT prediction that SVs in the sub-surface layers are energetically more favorable ~\citep{Gaberle2018}. 

\begin{figure*}
    \centering
    \includegraphics[width=\textwidth]{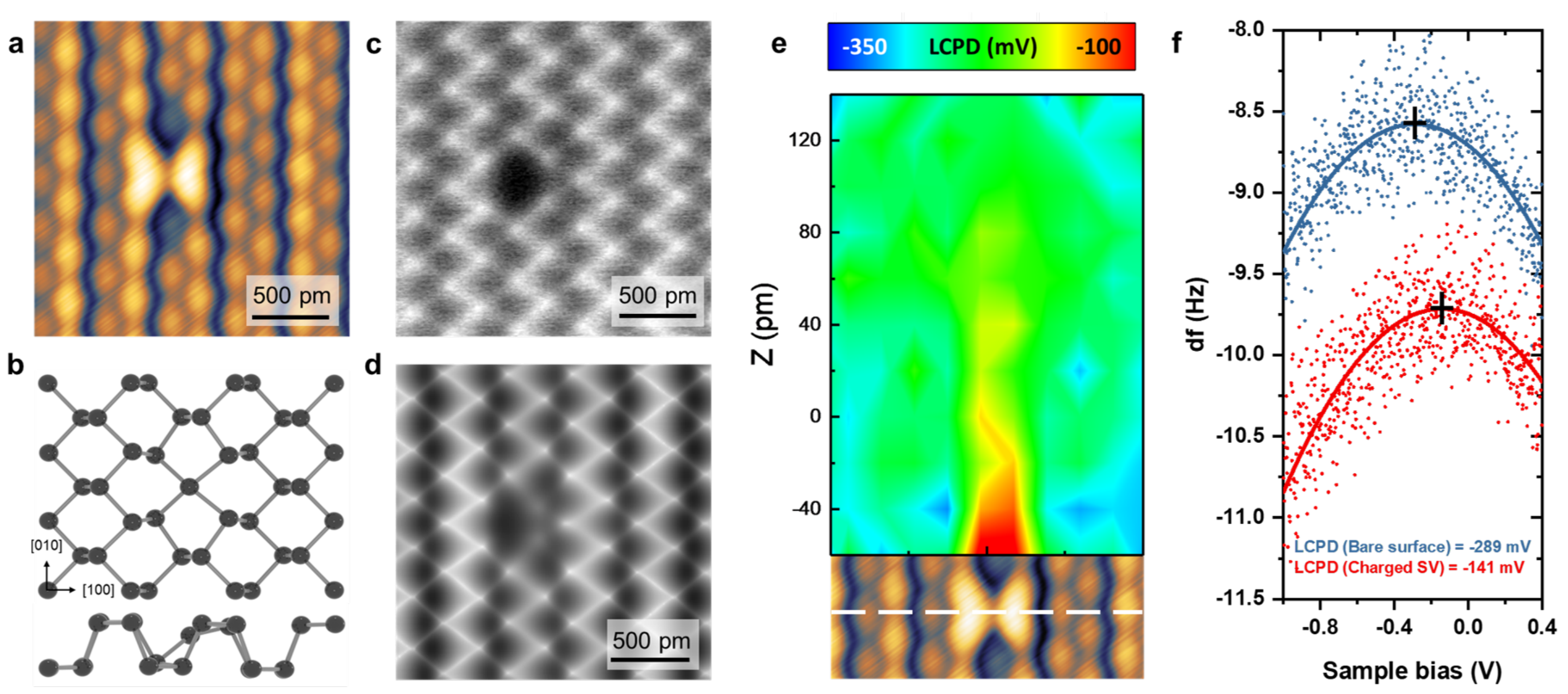}
    \caption{\textbf{Physical characterizations of $\mathbf{SV^-}$.}  (a) High resolution STM image of $\mathrm{SV}^-$ ($V_S = -0.6$ V, $I = 0.3$ nA). (b) Top and side views of atomic structure for $\mathrm{SV}^-$ with crystallographic directions.(c) Atom-resolved nc-AFM image of $\mathrm{SV}^-$. (d) Simulated nc-AFM image of $\mathrm{SV}^-$ by the the Probe Particle Model. (e) $x$-$z$ color mapping of spatial-dependent LCPD extracted from KPFM measurements across $\mathrm{SV}^-$, along the line marked at the bottom of STM image. (f) Frequency shift (df) measured as a function of applied sample bias right over and 2.12 nm away from $\mathrm{SV}^-$ at a relative height of $Z = -40$ pm with respect to the reference point: $Z = 0$ pm ($V_S = -1.0$ V, $I = 50$ pA). Parabolic fits and corresponding parabolic maximum values are indicated in the plot.}
    \label{fig:Characterization}
\end{figure*}

Upon thermal annealing at 453 K followed by rapid cooling, the majority of these dumbbell-shaped SVs vanished. Instead, a new type of defect manifested as a much smaller protrusion predominates the surface (Fig.~\ref{fig:Generation}d). A close examination revealed that this new defect adopts a butterfly-shaped feature spanning over two nearest neighbouring ZZ chains (Fig.~\ref{fig:Characterization}a). Moreover, the identical STM contrast of these butterfly-shaped defects (BSDs) suggests that these surface feastures are due to the reconstruction and migration of SVs during thermal annealing.

%

\begin{figure*}
    \centering
    \includegraphics[width=\textwidth]{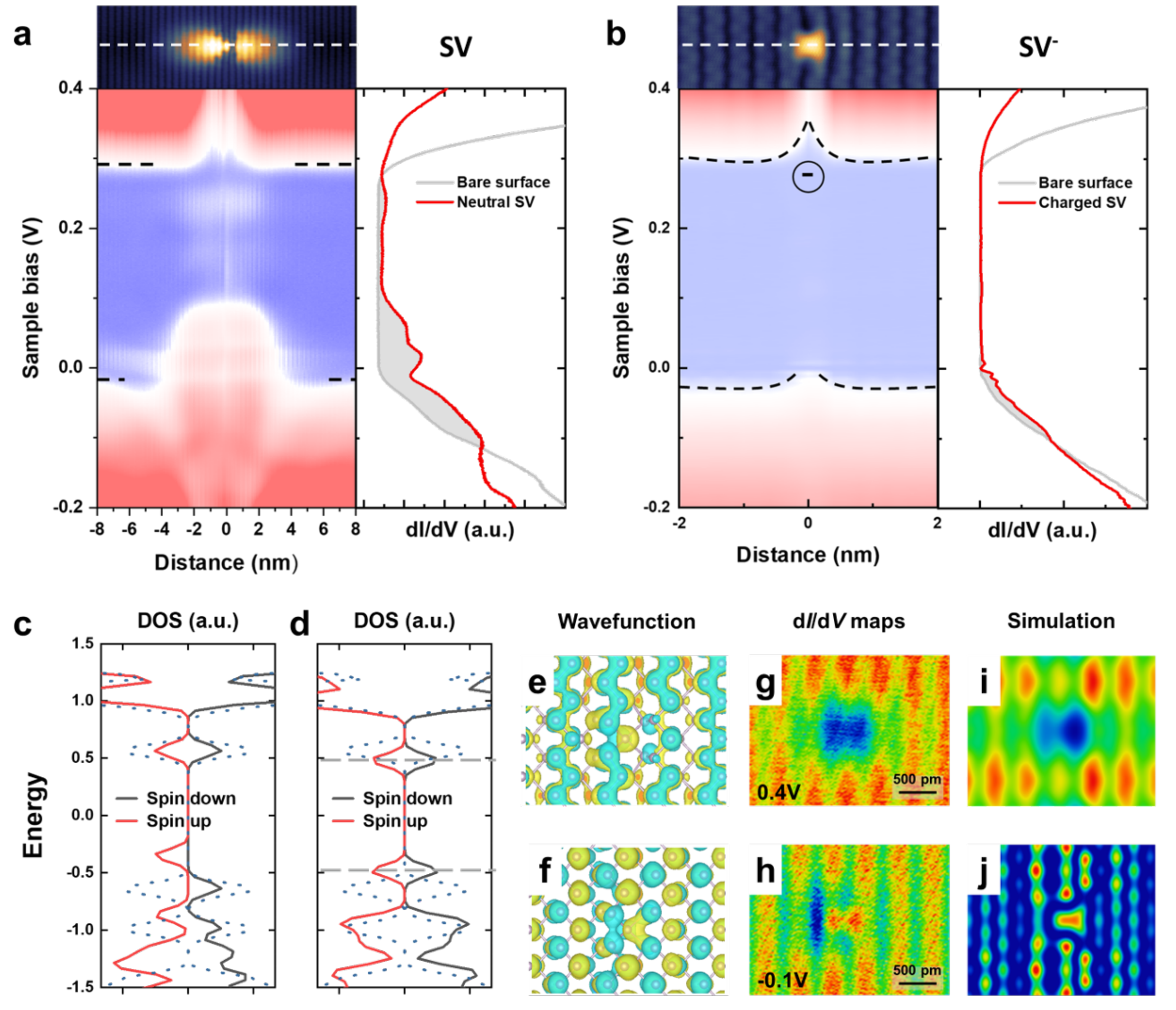}
    \caption{\textbf{Electronic structure of $\mathbf{SV^-}$.} (a, b) Color-coded $dI/dV$ spectra taken along AC direction across SV (a) and $\mathrm{SV}^-$ (b), where the positions for taken the line spectra are marked by white dashed lines in the corresponding STM images (upper panel). Point $dI/dV$ spectra taken above SV ($\mathrm{SV}^-$) is compared with the pristine surface in the right panels. (c, d) Calculated DOS for SV ($\mathrm{SV}^-$) and pristine BP (indicated by dash lines). (e, f) Wave functions of $\mathrm{SV}^-$ at energy levels close to CBM and VBM, which are indicated by grey dash lines in (d). (g, h) $dI/dV$ maps taken for $\mathrm{SV}^-$ at $V_S = 0.4$ V and $V_S = -0.1$ V. (i, j) Simulated $dI/dV$ maps at energy of $0.47$ V (i) and $-0.47$ V (j).}
    \label{fig:Probing}
\end{figure*}

Bias-dependent line $dI/dV$ spectra taken across the dumbbell and butterfly defects along the armchair (AC) direction (indicated by a white dashed line) are shown in the left panel of Fig.~\ref{fig:Probing}a and \ref{fig:Probing}b, respectively. A dumbbell-shaped SV mainly induces a strong LDOS modulation in the vicinity of the valence band maximum (VBM) as marked by grey color in the right panel of Fig.~\ref{fig:Probing}a. The in-gap states near VB of SV span $\approx 4$ nm away from the defect center, suggesting a shallow acceptor nature forming spatially extended bound hole states \citep{Qiu2017}. In contrast, BSD lacks in-gap states but exhibits an upward band bending of tens of meV for both VBM and CBM. In addition, $dI/dV$ maps taken at $0.4$ eV and $-0.1$ eV show a weak modulation, including a faint dark depression and protrusion over the defect site, respectively (Fig.~\ref{fig:Probing}g,h). All these observations consistently show that BSD is negatively charged.  

Kelvin probe force microscopy (KPFM) was further applied to characterize the local contact potential difference (LCPD) between BSD and bare BP surface. Fig.~\ref{fig:Characterization}f presents the KPFM results acquired right over and $\sim2.1$ nm away from BSD at a relative height of $Z = -40$ pm. A parabolic fitting of frequency shift $df-V$ curves reveals a positive shift of LCPD (i.e parabolic maximum) from $-289$ mV (over bare BP) to $-149$ mV (over BSD), 
conforming that BSDs are negatively charged ~\citep{Gross2009}. The LCPD ($x$, $z$) color-map acquired across BSD defect along the AC direction at different tip-sample distances further indicates a smaller LCPD and associated negative charge in close vicinity to BSD (Fig.~\ref{fig:Characterization}e), consistent with the calculated potential profile (Fig.~\ref{fig:Potential_Variation}).

To decipher the local structure of BSD, we performed nc-AFM imaging with a CO-functionalized tip (Fig.~\ref{fig:Characterization}c). The frequency shift image collected by nc-AFM reveals a dark hole-like feature surrounded by two bright spots and two dimmer spots. The protrusions in the nc-AFM image correspond to P atom positions due to their stronger repulsive interaction with CO-tip. Therefore, the nc-AFM image suggests BSD likely contains a missing top P atom (SV) or both top P and its bonded P atom at the bottom (divacancy denoted as DV). By DFT simulation of various defect structures via Probe Particle Model (See Fig. \ref{fig:Sym_AFM} in SI), the simulated nc-AFM (Fig.~\ref{fig:Characterization}d) of reconstructed SV reproduces the key features in experimental data well, including bright asymmetric protrusions at two sides of BSD. Therefore, BSD is determined to be negatively charged SV ($\mathrm{SV}^-$). The asymmetry in both STM and nc-AFM images of $\mathrm{SV}^-$ stems from the local reconstruction involving the bonding of central P atom with four adjacent P atoms (Fig.~\ref{fig:Characterization}b and Fig.~\ref{fig:Distance_analysis}), forming a hypervalent configuration with one extra negative charge.

The drastically different electronic properties of neutral SV and $\mathrm{SV}^-$ can be further understood by DFT calculations using hybrid pbe0 functional. Neutral SV shows the spin-polarized density of states (DOS) with in-gap states above the VBM (Fig.~\ref{fig:Probing}c). In contrast, the calculated DOS of $\mathrm{SV}^-$ resembles that of pristine BP (dashed blue curve) with a slight shift of both VB and CB towards each other (Fig.~\ref{fig:Probing}d). Upon the local reconstruction and ionization of SV, all the DBs are passivated in $\mathrm{SV}^-$, eliminating in-gap states and restoring the spin degeneracy in the DOS. Wavefunction plot of CB and VB band-edge states reveals a major contribution from dispersive band-like electronic states with small perturbation from the defect (Fig.~\ref{fig:Probing}e,f). In addition, the simulated $dI/dV$ maps at two energetic positions close to CB (Fig.~\ref{fig:Probing}i) and VB (Fig.~\ref{fig:Probing}j)side reproduce the key features in the experimental $dI/dV$ maps well. Therefore, a combination of DFT calculations and nc-AFM imaging unambiguously reveals that BSDs are $\mathrm{SV}^-$, presumably transformed from neutral SV.  

\begin{figure*}
    \centering
    \includegraphics[width=\textwidth]{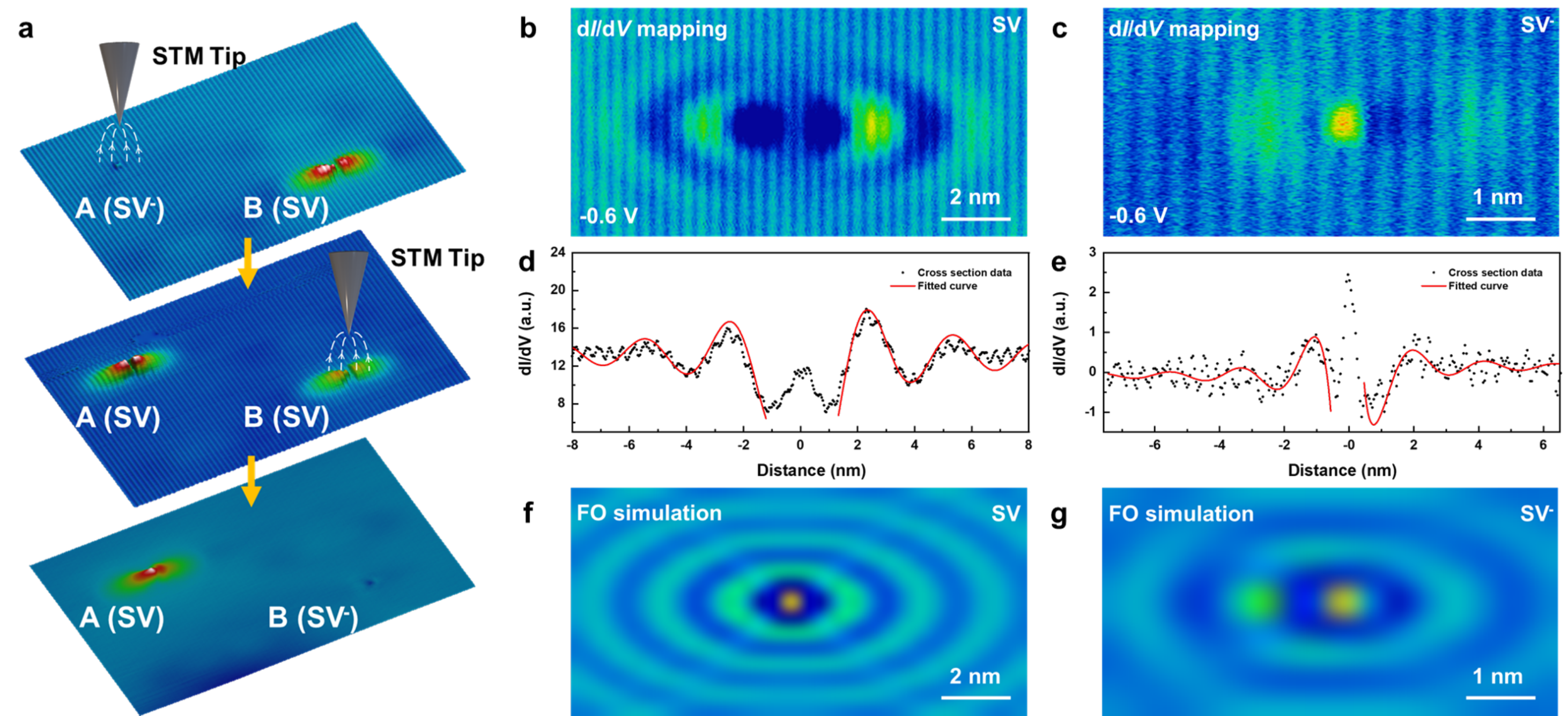}
    \caption{\textbf{Probing Friedel oscillations of SV and $\mathbf{SV^-}$.} (a) STM images of reversible transformation between $\mathrm{SV}^-$ and SV, controlled by tip manipulation. (b,c) spatial $dI/dV$ spatial maps taken at $V_S = -0.6$ V for SV and $\mathrm{SV}^-$ respectively. (d, e) Cross-section data across the center of SV and $\mathrm{SV^-}$ along the AC direction. The fitted curves for FO at both sides of defect are presented in red. Note that the discrepancy in the oscillation wavelength close to the defect center between experiment and theory can be attributed to a strong LDOS modulation at the VB edge of SV that may modify the local bulk bands. (f, g) Simulated spectral function maps at $-0.6$ eV for SV and $\mathrm{SV}^-$, respectively.}
    \label{fig:Friedel}
\end{figure*}

 In addition, we took advantage of STM tip manipulation to control the switch between SV and $\mathrm{SV}^-$ over two individual defects as shown in Fig.~\ref{fig:Friedel}a (from $\mathrm{SV}^-$ to SV for defect A and SV to $\mathrm{SV}^-$ for defect B). We note that a reversible switch over a single defect between SV and $\mathrm{SV}^-$ is challenging as high bias applied also may result in pulling nearby P atoms out from the surface, depending on the tip geometry. Nevertheless, positioning the STM tip over one charged defect A($\mathrm{SV}^-$) followed by a gradual increase of sample bias to $4.6$ V, triggers its switch back to neutral SV with the re-emergence of characteristic dumbbell shape  and in-gap states (Fig.~\ref{fig:dI_dV}). Similarly, upon applying a bias of $4.6$ V, another neutral B(SV) can be transformed into B($\mathrm{SV}^-$) with the appearance of butterfly shape and disappearance of in-gap states (Fig.~\ref{fig:dI_dV}). Such a tip-controlled transformation between SV and $\mathrm{SV}^-$ is presumably attributed to the ionization and de-ionization of defects induced by the local electric field or inelastic tunnelling process~\citep{Lee2010,Wong2015}. 

We then evaluated the impact of self-passivation of SV on the local charge scattering behavior by comparing the corresponding Friedel oscillations (FO) before and after ionization of SV. The scattering of charge carriers by defects often produces a periodic long-range modulation of the LDOS with periodicity associated with wave vector of scattered carriers ~\citep{Friedel1952}. We indeed observed contrasting LDOS oscillation pattern surrounding between SV and $\mathrm{SV}^-$. The characteristic energy-dependent wavelength of these oscillation patterns verifies their FO origin (Fig.~\ref{fig:Friedel}b-c and Fig.~\ref{fig:FO_dI_dV}). Neutral SV exhibits a highly anisotropic FO pattern with a stronger modulation along the AC direction, which can be attributed to a smaller effective carrier mass along this crystallographic axis ~\citep{Zou2016, Kiraly2019}. According to the standard FO theory ~\citep{Crommie1993a}, we performed the fitting of the line cross-section $dI/dV$ data for both SV and $\mathrm{SV}^-$ to the expression:  $A = A_0 \frac{\cos\left[\frac{2\pi\left(x - x_c\right)}{\lambda}+\phi\right]}{\left|x - x_c\right|^r} + C$, which yields a dramatic difference in the phase, amplitude and decay rate of FO patterns between SV and $\mathrm{SV}^-$. Firstly, the fitting of cross-section $dI/dV$ data to the equation above yields a phase difference between the oscillation at two sides of defect by $22.7^\circ$ for SV and $136.2^\circ$ for $\mathrm{SV}^-$. Hence, a small phase shift at SV produces a nearly symmetric FO pattern, while a significant phase shift results in an asymmetric FO pattern across $\mathrm{SV}^-$. Secondly, according to the fitting result, the oscillation amplitude in LDOS for SV ($A_L^0=8.59$, $A_R^0=11.14$) is nearly one order of magnitude larger than that for $\mathrm{SV}^-$ ($A_L^-=1.19$, $A_R^-=1.16$). Equally important, $\mathrm{SV}^-$ shows a faster decay rate ($-1.47$) than SV ($-0.96$), which may be the result of a weak perturbation and a large phase shift between two sides of $\mathrm{SV}^-$.

To understand the origin of different FO behaviours, we performed numerical calculations to simulate FO patterns for both SV and $\mathrm{SV}^-$ by treating BP as a single (valence) band 2D material with a direction-dependent carrier mass (See SI for details). Removing a P atom can be viewed as adding a P ``anti-nucleus." This negatively charged ``anti-nucleus" creates an attractive long-range potential for the holes, giving rise to FO and mid-gap bound states, as seen in Fig.~\ref{fig:Probing}a and c. Although the potential profile is non-trivial and must include screening effects, it is illuminating to get a rough estimate of the energy scales involved by assuming that the perturbation is limited to the unit cell with the missing P atom. First, we note that the lowest-energy mid-gap state for SV is located about $0.05$ eV above the VBM. Next, following the approach described in SI, we calculate the spectral function for the unit cell hosting the vacancy for a range of attractive potentials generated by the ``anti-nucleus," as shown in Fig.~\ref{fig:Bound_State_Pole}. From our analysis, we observe that the single-unit-cell potential of about $-1.5$ eV produces a localized state at the correct energy, which is thus set as the perturbation potential from SV to calculate the corresponding spectral maps of surrounding BP at different energies. The calculated spectral function map at the energy of $-0.6$ eV reveals the symmetric oscillation pattern, consistent with experimental $dI/dV$ maps of SV taken at the same energy (Fig.~\ref{fig:Friedel}b and f). The FO wavelength of $\approx 2$ nm in the AC direction obtained from the numerical simulations also agrees with the experimental results far from the defect (Fig.~\ref{fig:FO_comparison}). As for $\mathrm{SV}^-$, the local reconstruction involving the shift of central P atom to bond with adjacent four P atoms results in a more diffusive and asymmetric charge density, leading to a dipole-like potential splitting as shown in Fig.~\ref{fig:Potential_Variation}. To capture these effects, we split the perturbation ($-1.5$ eV) from SV into two unit cells. By adjusting the ratio and the separation (Fig.~\ref{fig:FO simulation}), we found that the calculated spectral function map with a split potential of $-1$ and $-0.5$ eV shows the best agreement with the experimental data (Fig.~\ref{fig:Friedel}c). In fact, the asymmetry and phase shift of spectral function maps is independent on the absolute value of split potentials (Fig.~\ref{fig:FO simulation}). This points out that making the potential less singular by spreading the perturbation out spatially leads to a reduced band distortion and carrier scattering. 

In summary, we demonstrated a new electronic self-passivation of SV in BP through the local reconstruction and ionisation of SV into negatively charged $\mathrm{SV}^-$ through a mild thermal annealing or STM tip manipulation, leading to the passivation of all the DBs and elimination of in-gap states. In contrast to SV with a strong and symmetric FO pattern, $\mathrm{SV}^-$ shows an asymmetric FO pattern with oscillation amplitude reduced by one order of magnitude and a faster decay rate, which can be attributed to its weak dipole-like perturbation. Our work opens up a new route for electronic self-passivation of defects, crucial for the further optimisation of the carrier mobility in BP and its analogues. 

\section{Acknowledgements}

J. Lu acknowledges the support from MOE grants (MOE2019-T2-2-044 and R-143-000-B58-114). M. Telychko acknowledges the support from A*STAR AME YIRG grant (Project No. A20E6c0098, R-143-000-B71-305). A.R. acknowledges the National Research Foundation, Prime Minister Office, Singapore, under its Medium Sized Centre Programme and the support by Yale-NUS College (through Grant No. R-607-265- 380-121). A.G. and P.J. acknowledge support of Czech Science Foundation project no. 20-13692X. 

%

\newpage

\onecolumngrid

\section{Supplementary Information}

\appendix
\renewcommand{\thefigure}{S\arabic{figure}}
\setcounter{figure}{0}

\section{Sample preparation} 
The bulk BP crystal (HQ graphene) was cleaved under ultrahigh vacuum to remove the degraded surface. Subsequently, the sample was heated up to 375 K to further remove potential absorbents before conducting the STM imaging and STS spectroscopic measurements at 4.8 K using Createc LT-STM system. The tungsten tip was calibrated spectroscopically to check the Shockley surface state on Au (111) surface. All the $dI/dV$ spectroscopy and mapping were obtained by a standard lock-in technique with a modulation voltage of 10mV and modulation frequency of 877Hz.

\section{DFT and Simulated SPM methods.}

\begin{figure}
    \centering
    \includegraphics[width = \textwidth]{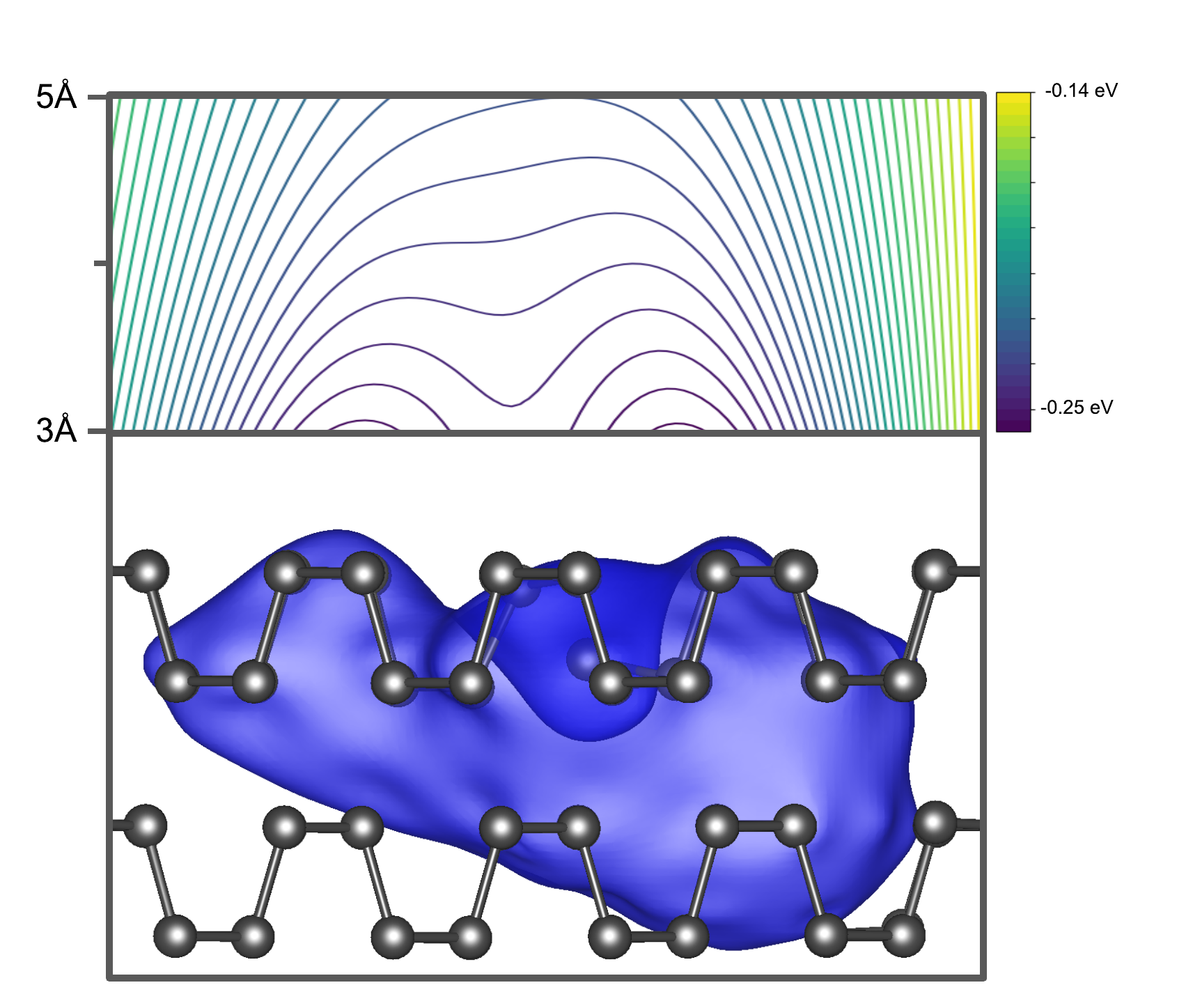}
  \caption{\textbf{2D and 3D plots of difference of hartree defined as $\delta{V}_h = {V}_h(q = -1) - {V}_h(q = 0)$.} The 2D plot is performed in a perpendicular plane that cuts the surface on the vacancy and extends through the armchair direction and the direction perpendicular to the surface. The 3D is an isosurface with value of $-0.35$ eV}
    \label{fig:Potential_Variation}
\end{figure}

All the studied structures were relaxed with the FHI-AIMS \citep{BLUM20092175} package using the hybrid pbe0 \citep{Ren_2012} potential with light wave functions and spin treatment, using the $\Gamma$-point to describe the reciprocal space. Van der Waals interaction was included using the Tkatchenko–Scheffler approximation \citep{PhysRevLett.102.073005} and we set a tolerance for the convergence of force and energy of $0.01$ eV/\AA\, and $10^{-5} $ eV, respectively. For the calculations, we used a 4-layers $4\times 4$ supercell of black phosphorus formed by 256 atoms (255 for the SV and 254 for the DV structures) of phosphorus, allowing the 3 top layers to relax with $20$ \AA\, of vacuum space above and lattice constants $a=3.31$ \AA\, and $b=4.38$ \AA\citep{doi:10.1063/1.1749671}.
After the simulations with AIMS, a $12\times 12$ supercell composed by 2303 atoms of the SV structure was built and studied for both neutral and charged state. The 5.25 nm (armchair) and 3.98 nm (zig-zag) distances between the vacancy and its periodic copies isolate it and allow us to study the large scale charge redistribution shown in Fig. \ref{fig:Potential_Variation}. To study this rather large system we used the home build code Fireball \citep{https://doi.org/10.1002/pssb.201147259} using the general gradient approximation BLYP \citep{PhysRevB.37.785} and vdW-D3\citep{doi:10.1063/1.3382344} with a local-orbital basis set with $R_c(s)=4.70$ a.u., $R_c(p)=5.20$ a.u. and $R_c(d)=5.0$ a.u.. To check the precision of the fireball calculations we replicated the calculation carried with AIMS of the $4\times 4$ supercell of the SV structure and obtained similar wavefunctions in shape and order.
The simulated AFM images were obtained with the probe particle model package \citep{PhysRevB.90.085421} using a stiffness of $k=0.25 \mathrm{Nm}^{-1}$ and a factor $q=-0.2$ for the electrostatic interaction included using the hartree potential of black phosphorus calculated with DFT.
Simulated $dI/dV$ images were obtained with the PP-STM code \citep{PhysRevB.95.045407}, using a s-like orbital as probe and the eigenstates from AIMS to describe the sample.

\section{Friedel oscillation fitting}
The cross-section data of FO can be divided into three parts: FO pattern on the left, the LDOS modulation by defect state at center, and FO pattern on the right. These three parts can be fitted and understood respectively: the defect state follows the Gaussian distribution along AC direction; the FO patterns on both sides are fitted by the expression for LDOS oscillations as a function of distance $x$:~\citep{Crommie1993a}

\begin{equation}
    A = A_0 \frac{\cos\left[\frac{2\pi\left|x - x_c\right|}{\lambda}+\phi\right]}{\left|x - x_c\right|^r} + C\,,
\end{equation}
where $A_0$ is the initial amplitude, $x$ is the independent variable, $x_c$ is the center position determined by extracting the peak center at defect state, $\lambda$ is the wavelength determined from the FFT results, $\phi$ is the phase of FO, $r$ is the decay rate determined by fitting the peak and valley intensity to distance in logarithm scale, $C$ is a constant value from the background.

\section{Analytical Formalism} 
To describe the pristine BP with added local potential, we use the single-band nearly-free electron model with a direction-dependent dispersion $\varepsilon_\mathbf{q} = \frac{\hbar^2}{2m_e}(\frac{q_x^2}{m^*_x}+\frac{q_y^2}{m^*_y})$, where $m_e$ is the electron mass, while $m_x$ and $m_y$ are the effective masses in arm-chair and zigzag directions. Note that 0.15$m_e$ in the AC direction and 1.18$m_e$ in the ZZ direction, where $m_e$ is the electron rest mass.~\citep{Qiao2014}). the single-band model contains one orbital per unit cell, setting the limit for the finest resolution scale.

The second-quantized Hamiltonian for the system can be written as
\begin{equation}
    \hat{\mathcal{H}} = 
    \sum_{\mathbf{q}} c^\dagger_\mathbf{q} \varepsilon_{\mathbf{q}}
    c_{\mathbf{q}}
    +
    \sum_\mathbf{r} c_{\mathbf{r}}^\dagger V_\mathbf{r}c_{\mathbf{r}} 
    =    c^\dagger_\mathbf{Q} H_\mathbf{Q}
    c_{\mathbf{Q}}
    +
   c_{\mathbf{R}}^\dagger \mathbf{V} c_{\mathbf{R}}\,.
\label{eqn:hamiltonian}
\end{equation}
Here, $c^\dagger_\mathbf{q}$ ($c_\mathbf{q}$) are fermionic creation (annihilation) operators for the single-band eigenstates, $c^\dagger_\mathbf{r}$ ($c_\mathbf{r}$) are their real-space counterparts operating on the unit cell at $\mathbf{r}$, and $V_\mathbf{r}$ is the position-dependent potential. To write the summation as matrix product, we combine $c_\mathbf{r}$ and $c_\mathbf{q}$ into vectors of operators $c_\mathbf{R}$ and  $c_\mathbf{Q}$, respectively. $H_\mathbf{Q}$ ($\mathbf{V}$) is a diagonal matrix with $\varepsilon_\mathbf{q}$ ($V_\mathbf{r}$) on the diagonal.

Using the fact that $c_\mathbf{q}$ and $c_\mathbf{r}$ are related by the Fourier transform, we have $c_\mathbf{R} = \Theta c_\mathbf{Q}$, where the elements of the matrix $\Theta$ are given by $\Theta_{jk} = \frac{1}{\sqrt{N}} e^{i\mathbf{r}_j\cdot \mathbf{q}_k}$ and $N$ is the number of unit cells in the system. With this, the Hamiltonian can be rewritten using the real-space operators as
\begin{equation}
    \hat{\mathcal{H}} = 
   c^\dagger_\mathbf{R}\left(
   \Theta 
   H_\mathbf{Q}
   \Theta^\dagger
    +
    \mathbf{V}
   \right) c_{\mathbf{R}}
   \,.
    \label{eqn:compact_hamiltonian}
\end{equation}

Next, we use this Hamiltonian to obtain the real-space Matsubara Green's function by subtracting the matrix in Eq.~\eqref{eqn:compact_hamiltonian} from $i\omega_n + \mu$ and taking the inverse:
\begin{equation}
    G_{i\omega_n + \mu} = \left(i\omega_n + \mu - \Theta 
   H_\mathbf{Q}
   \Theta^\dagger
    -
    \mathbf{V}
   \right)^{-1}
   = \left(\Xi_{i\omega_n + \mu}^{-1}
    -
    \mathbf{V}
   \right)^{-1}
   =
   \Xi_{i\omega_n + \mu}
   +
   \Xi_{i\omega_n + \mu}\mathbf{V}
   \left(1
    -
    \Xi_{i\omega_n + \mu}\mathbf{V}
   \right)^{-1}\Xi_{i\omega_n + \mu}\,.
\end{equation}
Here, $\Xi_{i\omega_n + \mu}$ is the real-space Green's function for the unperturbed system and the last term in the expression above is the correction due to $V_\mathbf{r}$.

We can obtain the spectral function by taking the diagonal elements of the Green's function $G_{i\omega_n + \mu}$ and substituting $i\omega_n\rightarrow$ with  $\omega + i0^+$ 
\begin{equation}
    \mathcal{A}_\omega\left(\mathbf{r}_i\right) =-2\mathrm{Im}\left[
    \Xi_{\omega + \mu}+ \Xi_{\omega + \mu}\mathbf{V} \left(1 - \Xi_{\omega + \mu}\mathbf{V}\right)^{-1}\Xi_{\omega + \mu}
    \right]_{ii}
    =-2\mathrm{Im}\left[
    \Xi_{\omega + \mu}^{ii}
    +
    \sum_{klm}
   \Xi_{\omega + \mu}^{ik}\mathbf{V}_{kl} \left[1 - \Xi_{\omega + \mu}\mathbf{V}\right]^{-1}_{lm} \Xi_{\omega + \mu}^{mi}
    \right]
    \,,
    \label{eqn:A}
\end{equation}
where $\omega$ is measured relative to the Fermi level. Given that $\mathbf{V}$ is diagonal and only contains non-zero entries for the unit cells containing the local potential, the expression inside the brackets simplifies to
\begin{align}
   & 
    \Xi_{\omega + \mu}^{ii}
    +
    \begin{pmatrix}
        \Xi_{\omega + \mu}^{i1} & \Xi_{\omega + \mu}^{i2}&\dots
    \end{pmatrix}
   \tilde{\mathbf{V}} \left(1 - \tilde{\Xi}_{\omega + \mu}\tilde{\mathbf{V}}\right)^{-1}
    \begin{pmatrix}
        \Xi_{\omega + \mu}^{1i} \\ \Xi_{\omega + \mu}^{2i}\\\vdots
    \end{pmatrix}\,,
\end{align}
where $\mathbf{\tilde{V}}$ only contains the unit cells where local potential is added and $\tilde{\Xi}_{\omega + \mu}$ is the portion of $\Xi_{\omega + \mu}$ including only these unit cells.

To calculate $\Xi_z$, first note that $\Theta$ is a unitary matrix so that
\begin{equation}
    \Xi_z = 
     \left(z - \Theta 
   H_\mathbf{Q}
   \Theta^\dagger
   \right)^{-1} =
   \left(\Theta z \Theta^\dagger- \Theta 
   H_\mathbf{Q}
   \Theta^\dagger
   \right)^{-1}
    =
  \Theta \left( z - 
   H_\mathbf{Q}
   \right)^{-1}\Theta^\dagger\,.
\end{equation}
The matrix elements of $\Xi_z$ are given by
\begin{equation}
    \Xi^{ik}_z =\sum_{lm} \Theta_{il} \left(z - H_\mathbf{Q}\right)_{lm}^{-1}\Theta^\dagger_{mk}
    =\sum_{l} \Theta_{il} \left(z - H_\mathbf{Q}\right)_{ll}^{-1}\Theta^\dagger_{lk}
      = \frac{1}{N}\sum_\mathbf{q} \frac{e^{i\mathbf{q}\cdot\left(\mathbf{r}_i - \mathbf{r}_k\right)}}{z - \varepsilon_\mathbf{q}}\,,
\end{equation}
where we used the fact that $H_\mathbf{Q}$ is a diagonal matrix. Changing the momentum sum into an integral yields
\begin{align}
    \Xi_z^\mathbf{r} 
    &= 
    \frac{1}{N}\frac{A}{\left(2\pi\right)^2}\int d^2\mathbf{q}
    \frac{e^{i\mathbf{q}\cdot \mathbf{r}}}{z - \frac{\hbar^2}{2m_e}\left(\frac{q_x^2}{m_x}+\frac{q_y^2}{m_y}\right)}=  \frac{\mathcal{V}}{2\pi}\frac{2m_e\sqrt{m_xm_y}}{\hbar^2}
    \int dp\, p
    \frac{ J_0\left(p\tilde{r}\right)}{z - p^2}
    \,,
\end{align}
where $A$ is the area of the system, $\mathcal{V} = A / N$ is the area of the unit cell, and  $\tilde{r} = \sqrt{\frac{2 m_e}{\hbar^2} \left( m_x x^2 + m_y y^2 \right)}$.

For $\mathbf{r}\neq 0$,
\begin{equation}
\label{eqn:xi_r_not_zero}
    \Xi_z^{\mathbf{r}\neq 0} 
    =
    \frac{\mathcal{V}}{2\pi}\frac{2m_e\sqrt{m_x m_y}}{\hbar^2}
    \int_0^\infty dp\, p
    \frac{J_0 \left(p\tilde{r}\right)}{z - p^2}
    =
    -\frac{\mathcal{V}}{2\pi}\frac{2m_e\sqrt{m_x m_y}}{\hbar^2}
    K_0\left(\tilde{r}\sqrt{-z}\right)\,,
\end{equation}
where $K_0$ is the modified Bessel function. When $\tilde{r} = 0$, we need to introduce a cutoff $\sqrt{C}$ so that the integral does not diverge
\begin{equation}
\label{eqn:xi_r_zero}
    \Xi_z^{\mathbf{r}= 0} 
    =
   \frac{\mathcal{V}}{2\pi}\frac{2m_e\sqrt{m_x m_y}}{\hbar^2}
    \int_0^{\sqrt{C}} dp
    \frac{p}{z - p^2} =-\frac{\mathcal{V}}{2\pi}\frac{2m_e\sqrt{m_x m_y}}{\hbar^2}\frac{1}{2}\ln\left(1-\frac{C}{z}\right) \, .
\end{equation}
We now consider the choice of $C$. Recall that $-2\mathrm{Im}\left[\Xi^{\mathrm{r} = 0}_{\omega}\right]$ is the spectral function for the pristine system whose integral over the energies $\omega \in (-\infty, \infty)$, yields $2 \pi$. Using this fact, we have
\begin{equation}
\label{eqn:complex_log_integral}
    \int^{\infty}_{-\infty} d\omega \, \textrm{Im} \left[  \frac{\mathcal{V}}{\pi}\frac{m_e\sqrt{m_x m_y}}{\hbar^2}\ln\left(1-\frac{C}{\omega + i0^+}\right)  \right] = \mathcal{V}\frac{m_e\sqrt{m_x m_y}}{\hbar^2} C = 2 \pi \,,
\end{equation}
leading to
\begin{equation}
    C \rightarrow \frac{2\hbar^2 \pi}{\mathcal{V} m_e \sqrt{m_x  m_y }} = \frac{4\hbar^2 \pi a_0^2}{2\mathcal{V} a_0^2 m_e \sqrt{m_x  m_y}} = \frac{4 \pi}{\tilde{\mathcal{V}} \sqrt{m_x m_y}}\mathrm{Ry} \, ,
\end{equation}
where $\tilde{\mathcal{V}} = \frac{\mathcal{V}}{a_0^2}$, $\mathrm{Ry}= \frac{\hbar^2}{2 m_e a_0^2}$ is the Rydberg energy, and $a_0$ is the Bohr radius. Finally, $\Xi_z^{\mathbf{r}}$ is given by 

\begin{align}
    \Xi_z^{\mathbf{r}\neq 0} 
    &=
    -\frac{2}{C}
    K_0\left(\sqrt{m_x x^2 + m_y y^2}\sqrt{-\frac{z}{\mathrm{Ry}}}\right)
    \,, \label{eqn:prop_r_not_zero}
    \\
    \Xi_z^{\mathbf{r}= 0} 
    &=
    -\frac{1}{C} \ln\left(1-\frac{C}{z}\right) \label{eqn:prop_r_zero} \,. 
\end{align}

All the field-theoretic numerical calculations were performed using Julia programming language.~\citep{Bezanson2017} The code is available at https://github.com/rodin-physics/black-phosphorus-defects.

\clearpage

\begin{figure}
    \centering
    \includegraphics[width = 0.7\textwidth]{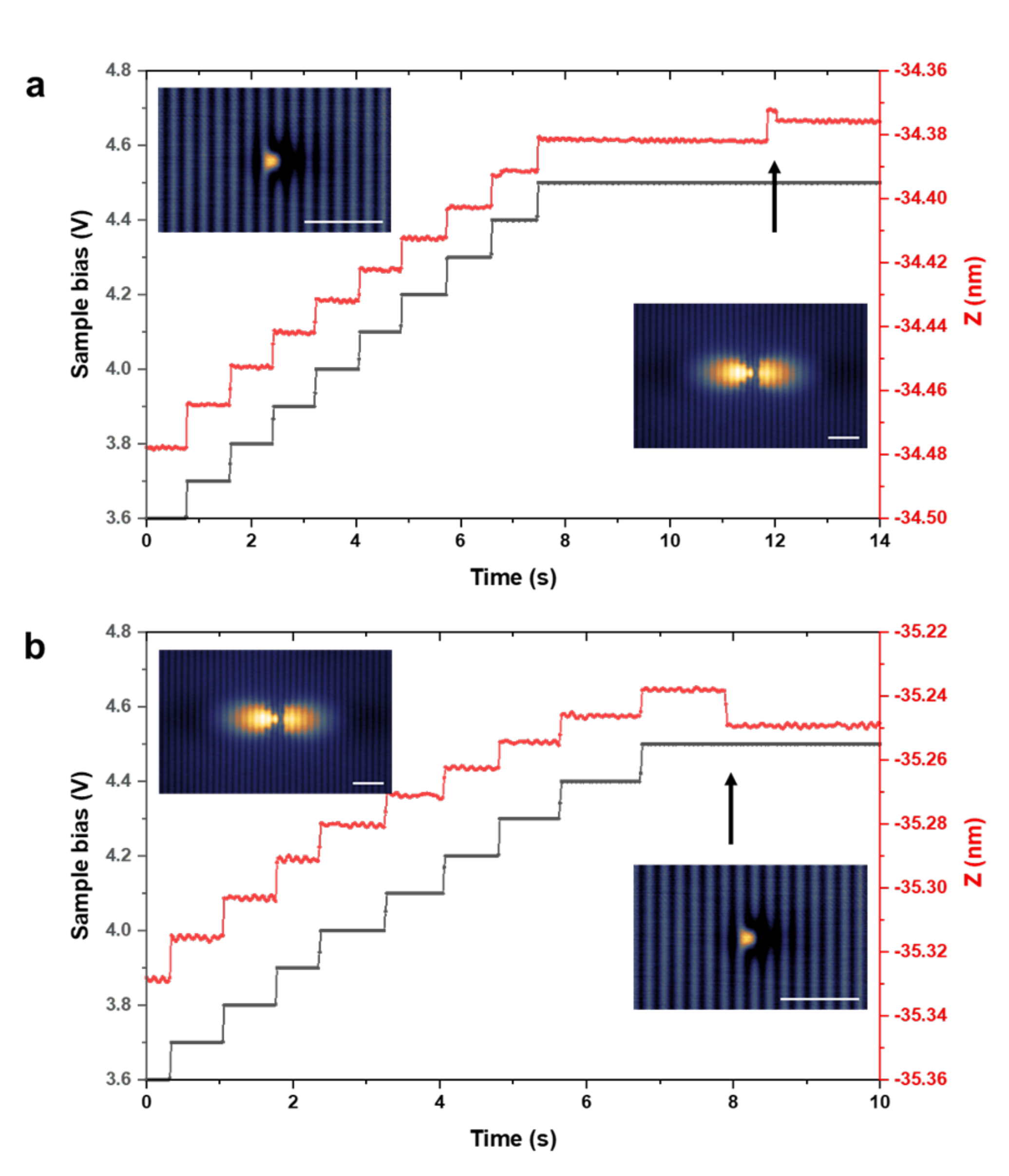}
    \caption{\textbf{Real-time monitoring the ionization and de-ionization of single vacancy in BP.} Recorded sample bias and tip height for the transformation (a) from $\mathrm{SV}^-$ to SV and (b) SV to $\mathrm{SV}^-$. A sudden jump (drop) in the tip height channel is marked by the arrow.}
    \label{fig:Switching}
\end{figure}
\clearpage
\begin{figure}
    \centering
    \includegraphics[width = \textwidth]{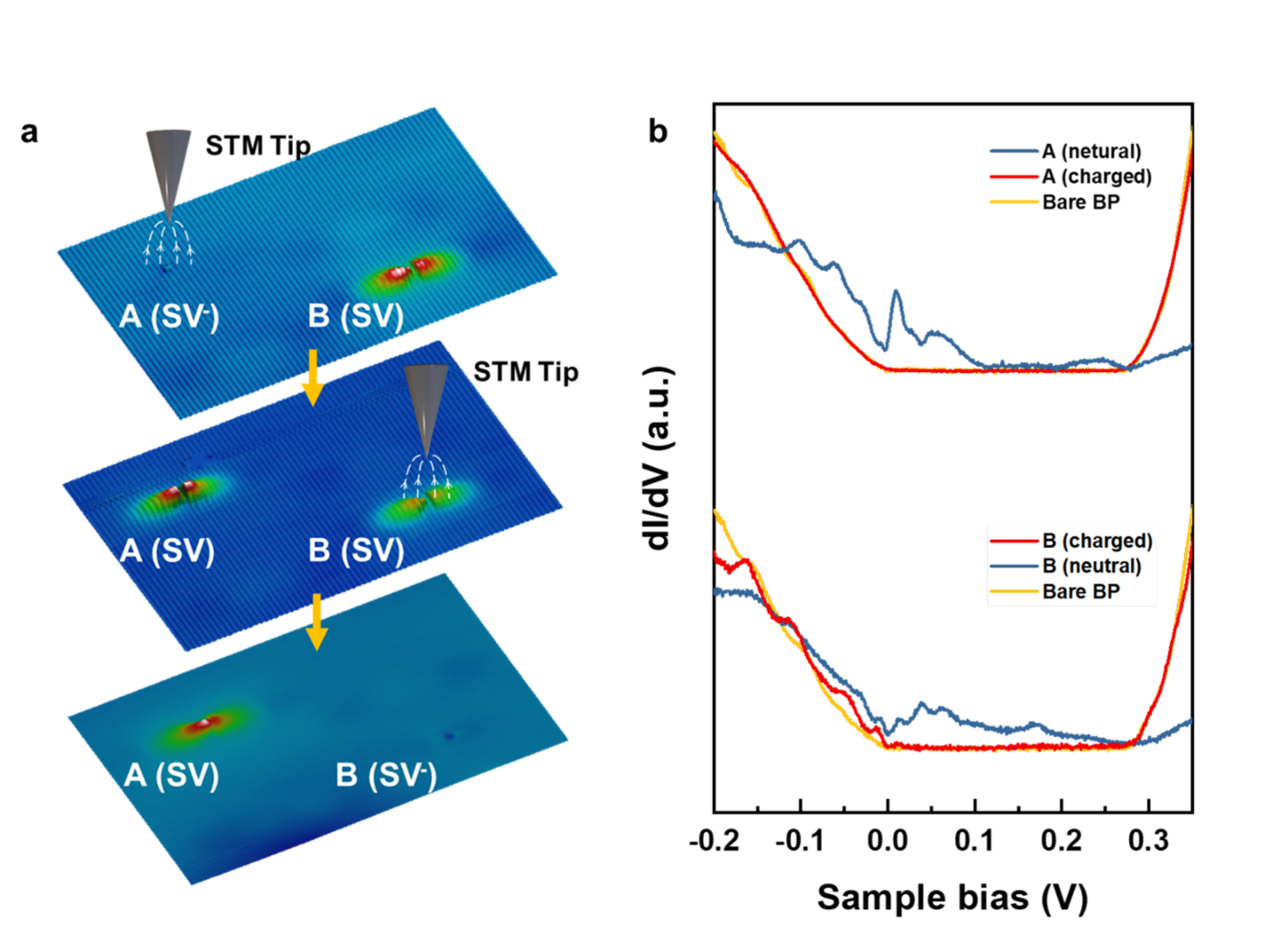}
    \caption{\textbf{STM imaging and $dI/dV$ measurements of single vacancy A and B during their transformation process.} (a) The illustration of the step-by-step transformation as introduced in the main text. (b) $dI/dV$ spectra taken at SV(A) and SV(B) before and after ionization. The reference STS curve taken at the bare surface is plotted in yellow }
    \label{fig:dI_dV}
\end{figure}
\begin{figure}
    \centering
    \includegraphics[width = \textwidth]{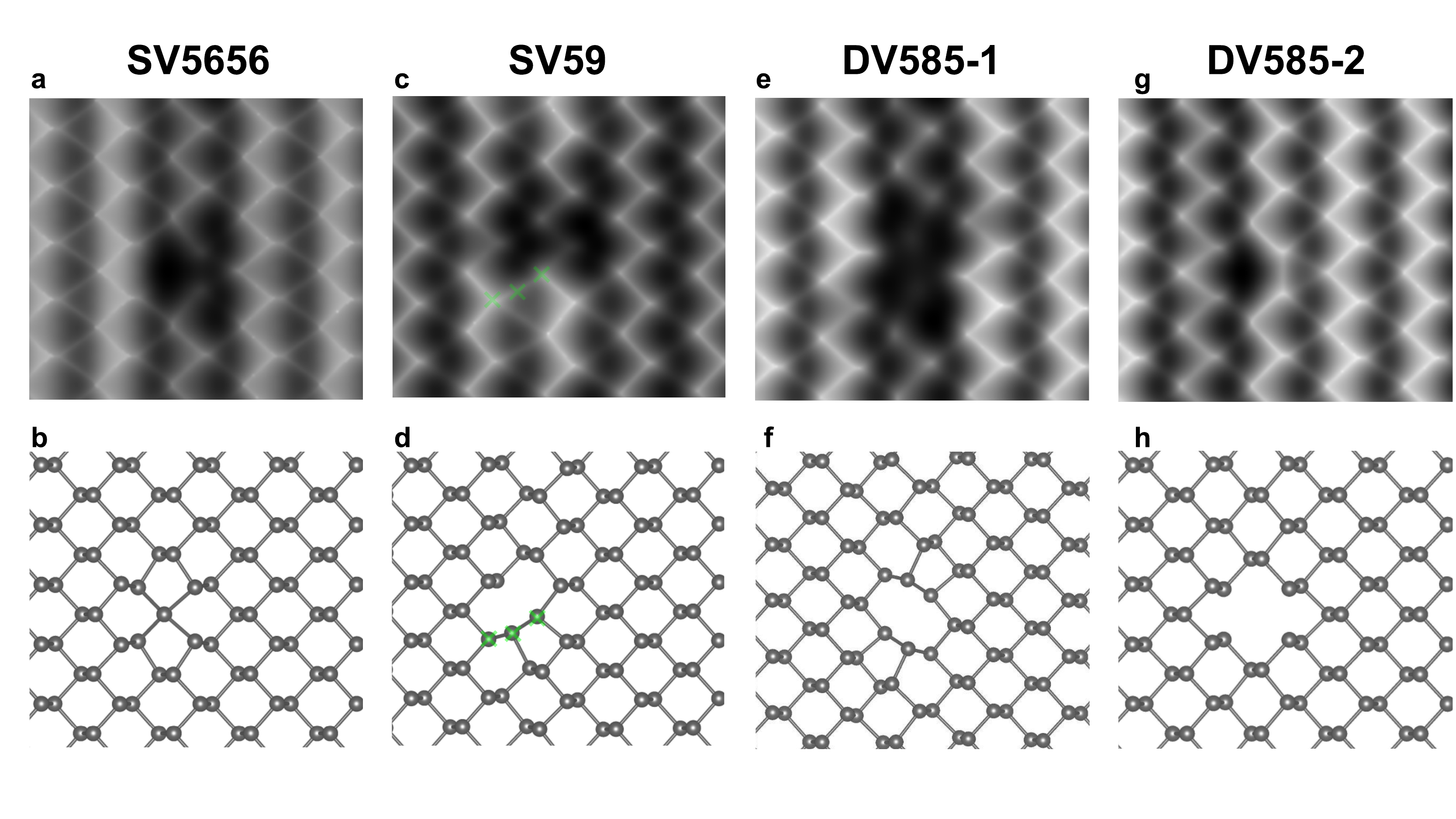}
    \caption{\textbf{Simulated nc-AFM images of various vacancy structures in BP by Probe Particle Model} (a) single vacancy with 2 five-member rings and 2 six-member ring at defect site (SV5656), (c) single vacancy with 1 five-member ring and 1 nice-member ring at defect site (SV59), (e) double vacancy with 2 five-member rings and 1 eight-member ring in asymmetric structure at defect site (DV585-1) and (g) double vacancy with 2 five-member rings and 1 eight-member ring in symmetric structure at defect site (DV585-2). The corresponding atomic structures of (a), (c), (e) and (g) are shown in (b), (d), (f) and (h) respectively, showing only the top layer for clarity. The positions of the asymmetric atoms in (c,d) were mark with green crosses for clarity.}
    \label{fig:Sym_AFM}
\end{figure}
\begin{figure}
    \centering
    \includegraphics[width = 0.7\textwidth]{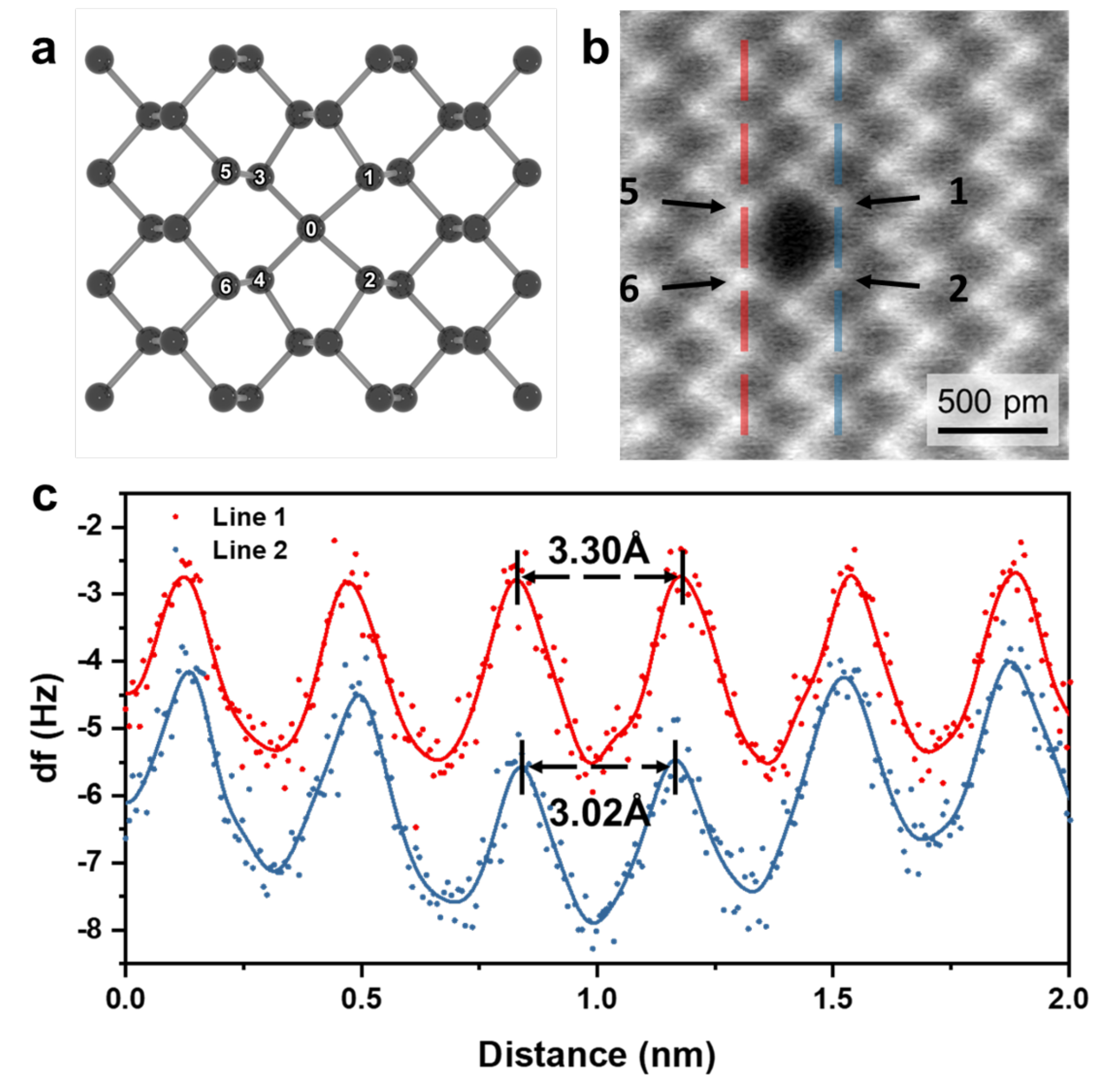}
    \caption{\textbf{Analysis of the structural relaxation of $\mathrm{SV}^-$ via nc-AFM imaging.} (a) Top view of atomic structure for $\mathrm{SV}^-$. The numbers are labelled for the P atoms at the defect site. (c) Cross-section data extracted along two dash lines in the nc-AFM image (b). The distance measured between the P(5,6)(Line 1) and P(1,2) (Line 2) are marked in the plot. 
    To better describe this, we first labeled the P atoms at the center as 0 and surrounding four P atoms as 1, 2, 3, 4, respectively. As illustrated in the top view, P atoms (1, 2) and P (3,4) are two nearby P atoms on the upper and lower side of the buckled surface layer, respectively. P atoms (5,6) are located at the upper side and bonded to P (3,4) at the bottom side. The local reconstruction involves the shift of P (0) in the original neutral SV towards the center to bond with P atoms (1, 2) with their DBs saturated. As a result,  P atoms (1, 2) are pulled down by the central P (0), resulting in a dimmer contrast, while P(5,6) are slightly buckled up, leading to a brighter contrast. Such a local reconstruction of SV also shortens the distance between P(1, 2) and P(5, 6) from $0.33$ nm to $0.30$ nm, while the distance between P(3,4) has negligible variation, consistent with our nc-AFM result.}
    \label{fig:Distance_analysis}
\end{figure}
\begin{figure}
    \centering
    \includegraphics[width = \textwidth]{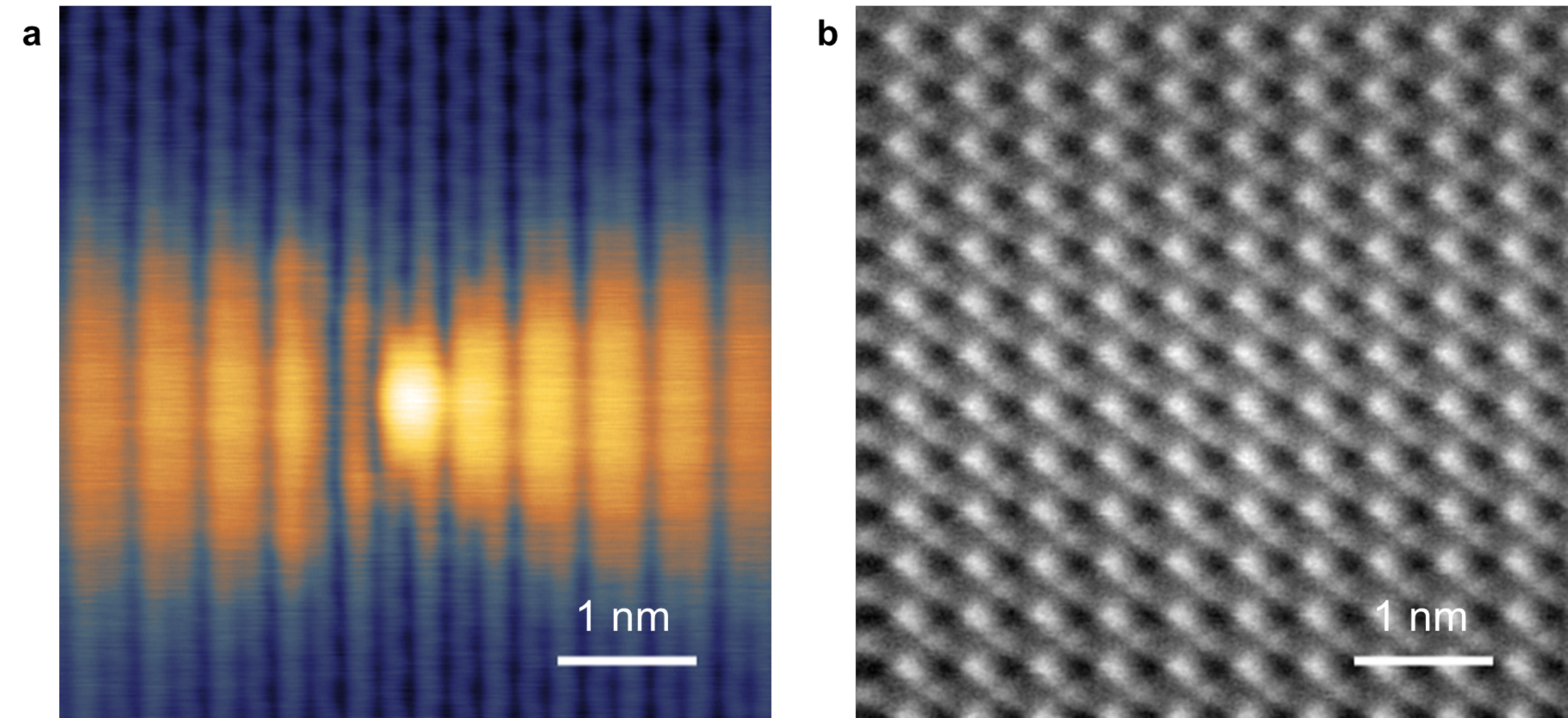}
    \caption{\textbf{Neutral SV resides in the sub-surface layer.} (a) STM image taken at $V_S = -50$ mV, $I = 50$ pA, revealing a typical dumbbell-shaped feature for neutral SV. (b) Constant-height nc-AFM image taken at the same region for neutral SV at relative height of $Z = 0$ pm with respect to the reference point ($V_S = -50$ mV, $I = 5$ pA) shows an intact BP lattice, indicating that the neutral SV is buried under the surface.}
    \label{fig:SV_UnderSurface}
\end{figure}
\begin{figure}
    \centering
    \includegraphics[width = \textwidth]{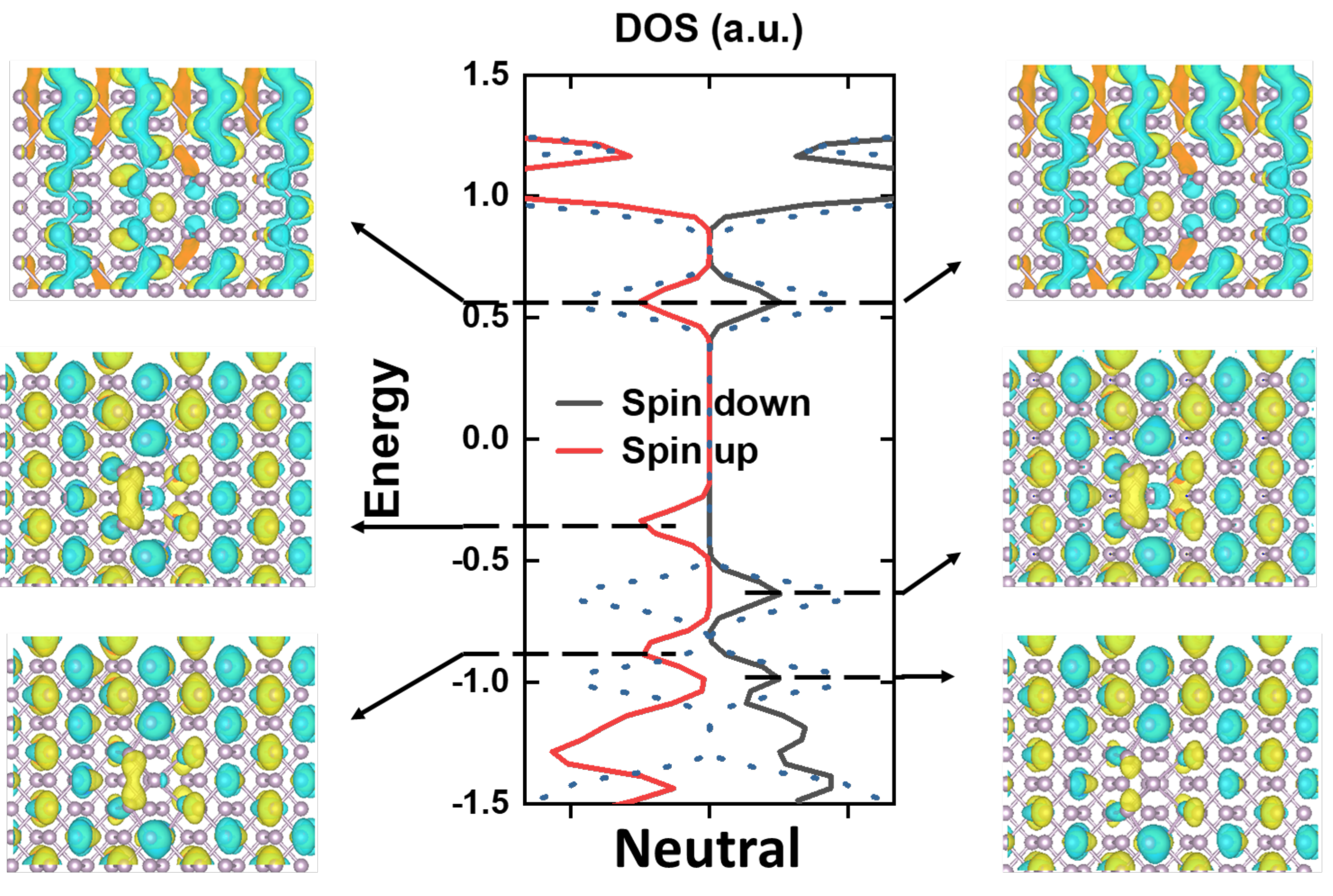}
    \caption{\textbf{Calculated DOS and the corresponding wave function plots of neutral SV at different energies.} Note the DOS of intrinsic BP is shown in dash line. }
    \label{fig:WF_neutral}
\end{figure}
\begin{figure}
    \centering
    \includegraphics[width = \textwidth]{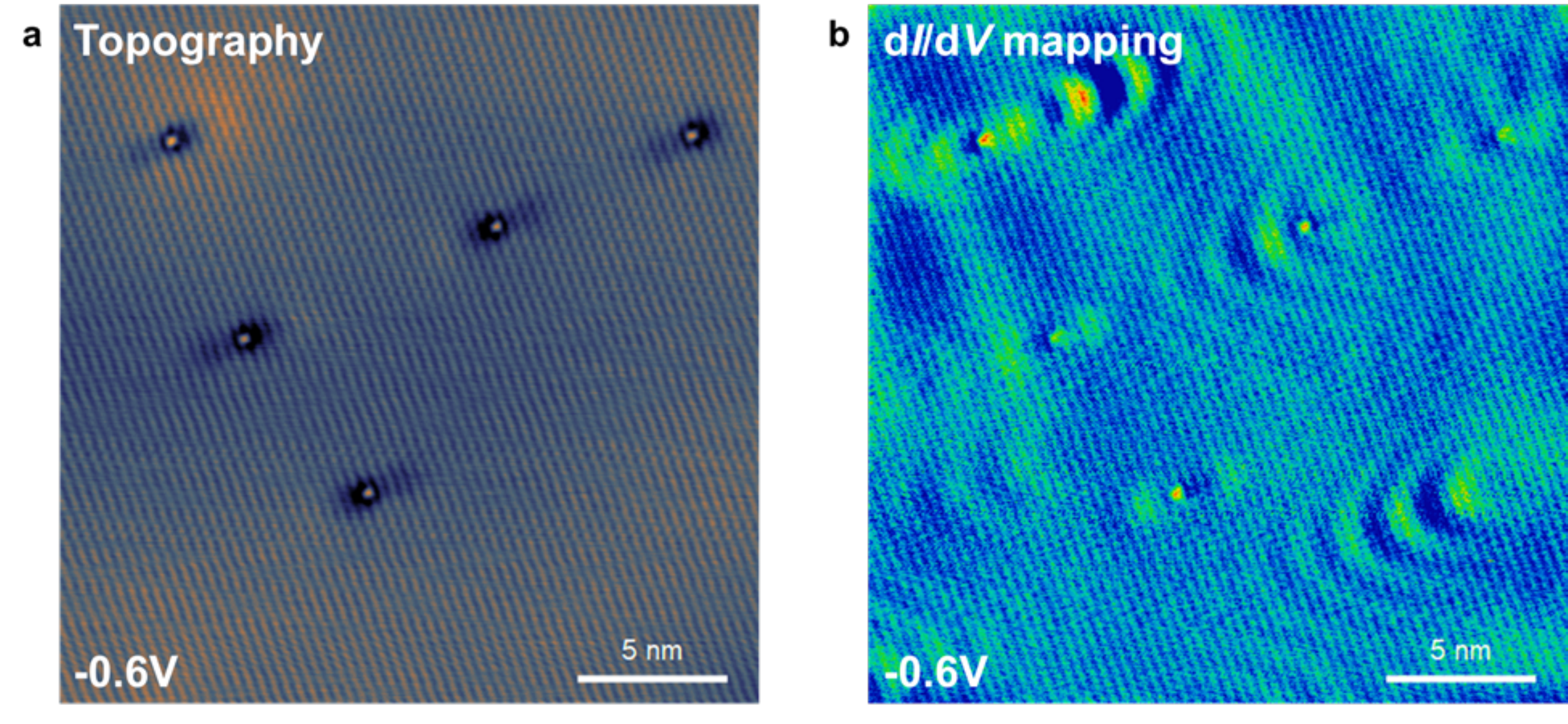}
    \caption{\textbf{Resolving FO patterns from neutral SV and charged SV in the same area.} (a) STM image reveals 5 $\mathrm{SV}^-$ on BP surface. (b) the corresponding $dI/dV$ map taken in the same region. In addition to the FO of $\mathrm{SV}^-$, intensive FO from neutral SV buried under the surface was observed}
    \label{fig:FO_comparison}
\end{figure}
\begin{figure}
    \centering
    \includegraphics[width = 0.7\textwidth]{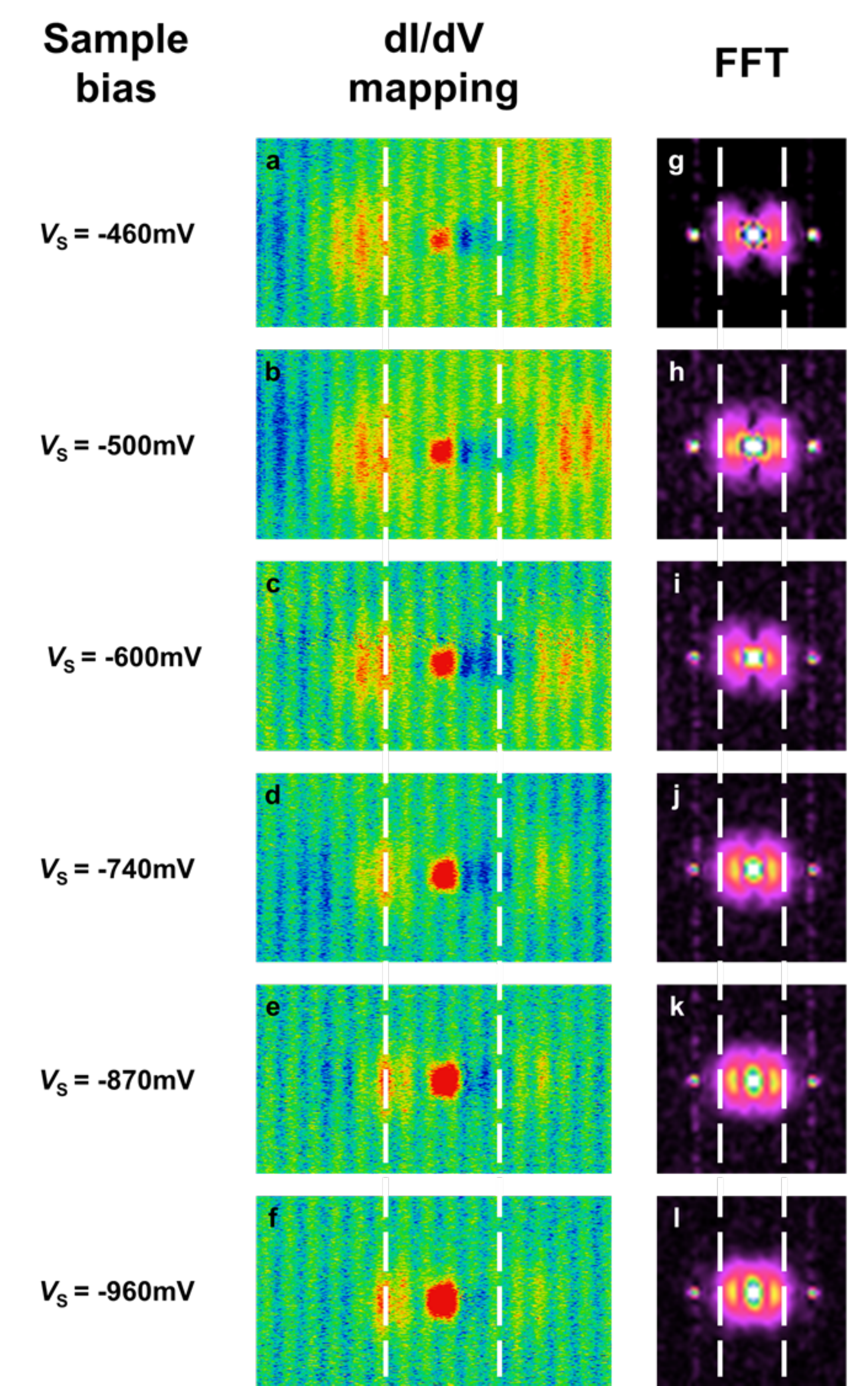}
    \caption{\textbf{Energy-dependent FO pattern of $\mathbf{SV^-}$.} (a-f) $dI/dV$ maps for $\mathrm{SV}^-$ at different sample bias. (g-l) Corresponding FFT images. The white dash lines are used to guide the eye. It reveals the change of the wavelength and corresponding reciprocal points as a function of sample bias.}
    \label{fig:FO_dI_dV}
\end{figure}
\begin{figure}
    \centering
    \includegraphics[width = \textwidth]{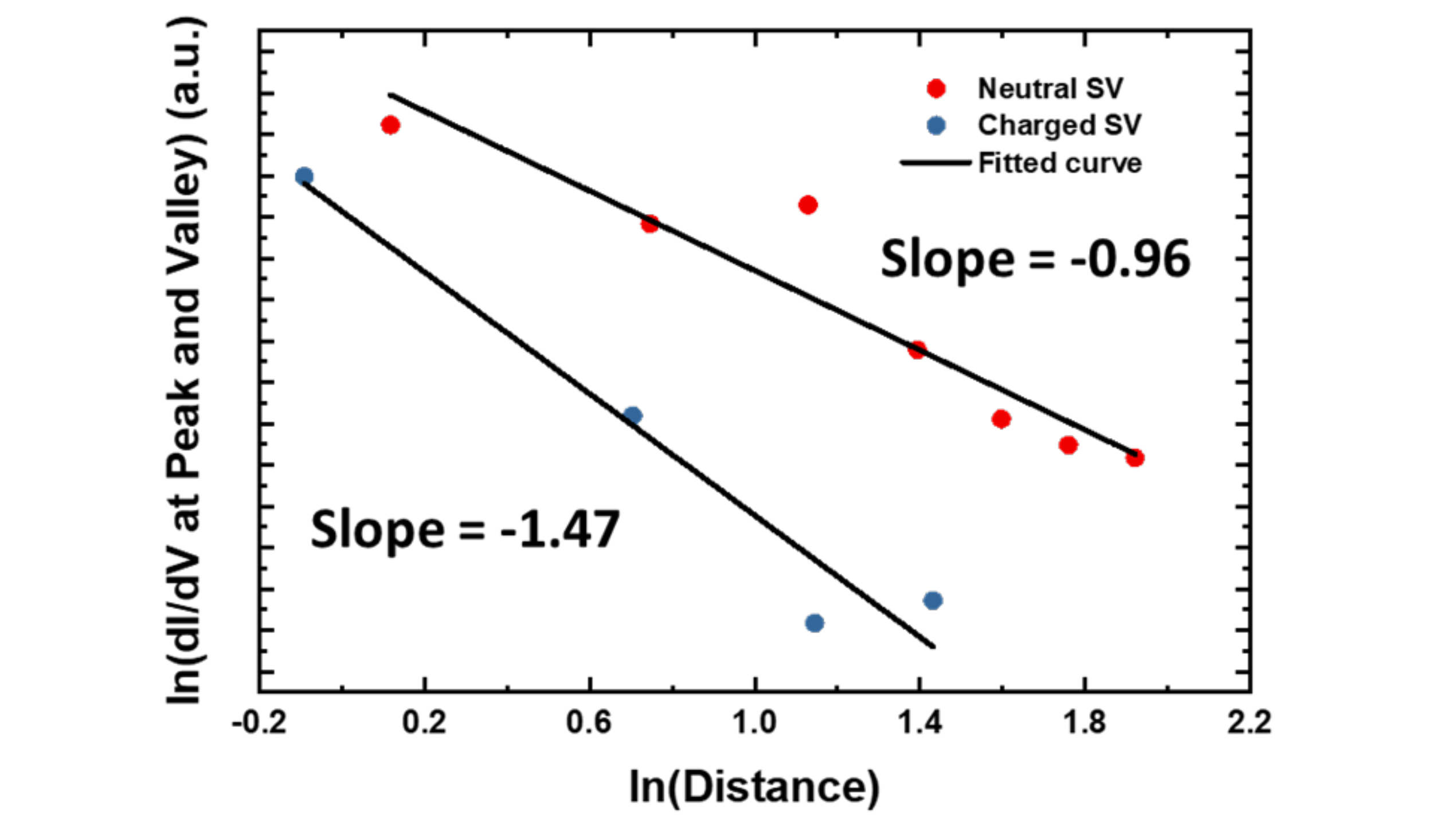}
    \caption{\textbf{Determination of the decay rate of FO at SV and $\mathbf{SV^-}$.} The peak and valley intensity as a function of positions is plotted in a logarithm scale. A linear fit is used to obtain the decay rate for SV and $\mathrm{SV}^-$. The derived slopes are indicated in the plot.}
    \label{fig:Decay_rate}
\end{figure}
\begin{figure}
    \centering
    \includegraphics[width = 0.7\textwidth]{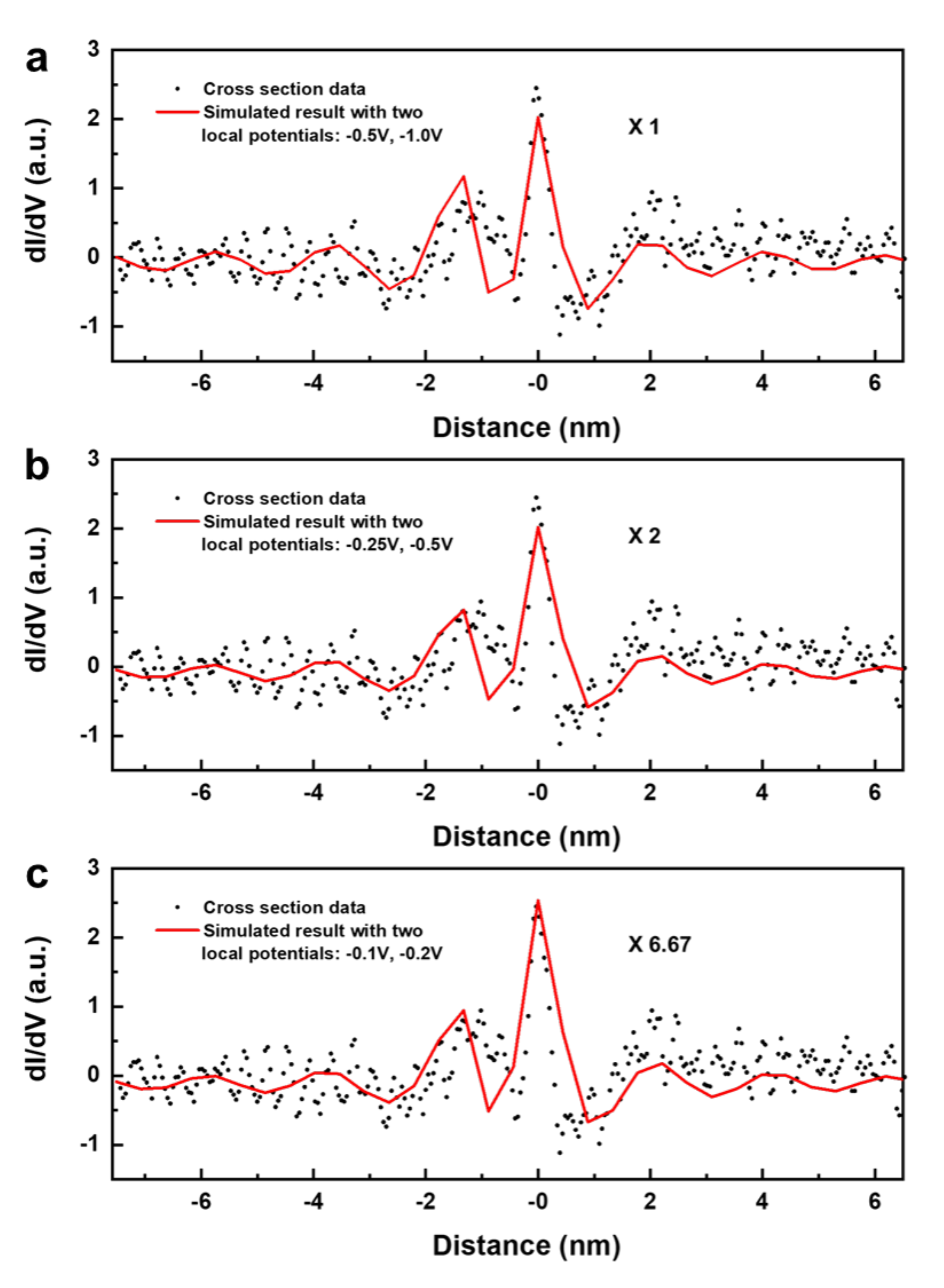}
    \caption{\textbf{Simulated d\textit{I}/d\textit{V} line profile of FO at $\mathrm{SV}^-$ } (a) d\textit{I}/d\textit{V} mapping for $\mathrm{SV}^-$ taken at $V = -0.6$ V. (b) Cross section data extracted from line cut in (a) with simulated results overlapped. The simulation is based on a two-charge centre separated at $\mathrm{SV}^-$}
    \label{fig:FO simulation}
\end{figure}

\begin{figure}
    \centering
    \includegraphics[width = 0.7\textwidth]{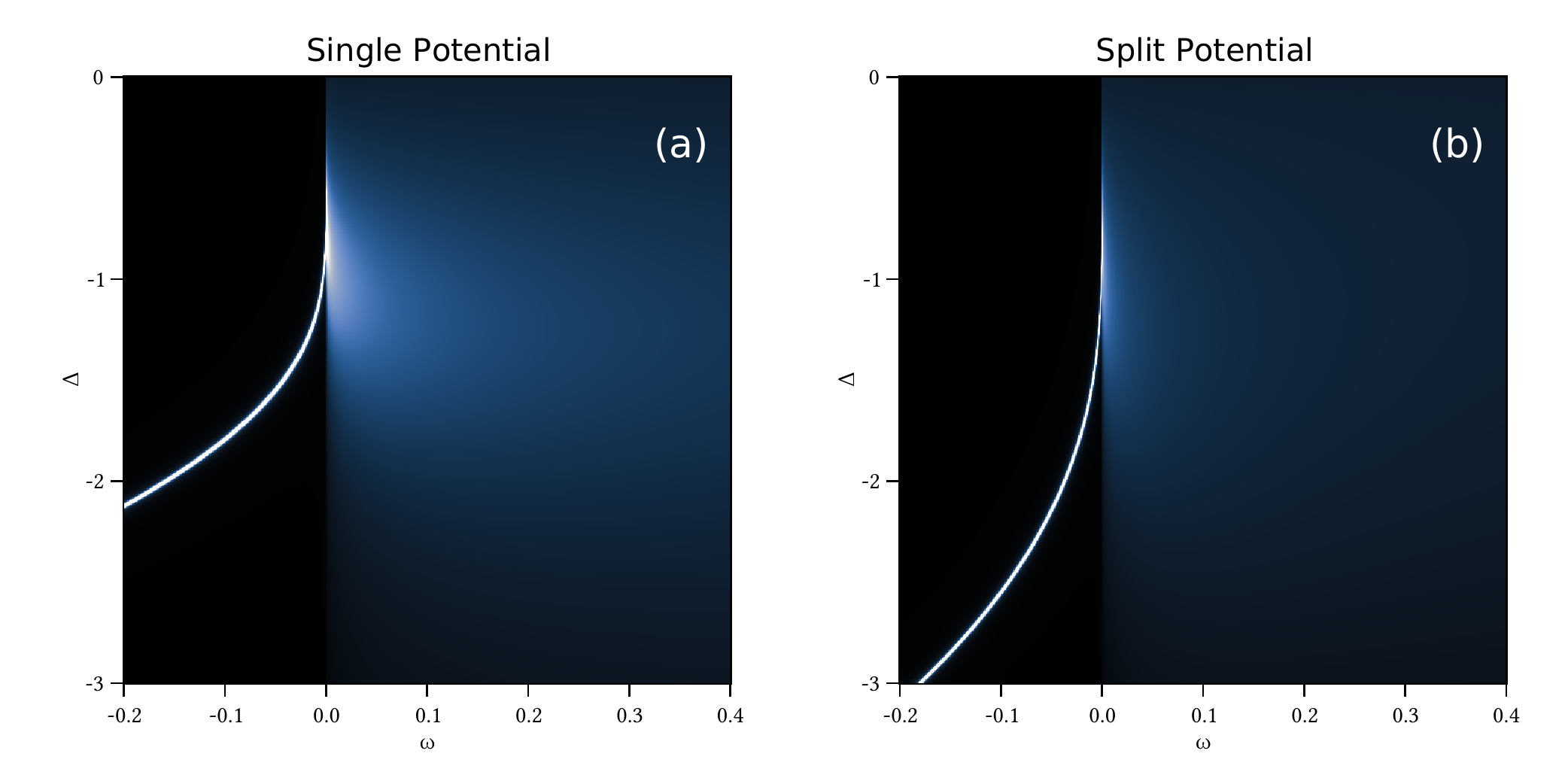}
    \caption{\textbf{Spectral function at the origin.} (a) Spectral function at the unit cell at the origin as a function of the local energy perturbation $\Delta$ also at the origin for a range of energies $\omega$. Positive $\omega$'s correspond to the band energies, while the white line gives the position of the mid-gap bound state. (b) Spectral function at the origin for a split potential. $\Delta$ is split into two portions so that the perturbation in the unit cell immediately to the right of the origin is $\Delta/ 3$, while the perturbation two unit cells to the left of the origin is $2\Delta / 3$. Note the much shallower energies of the bound state for the split potential.}
    \label{fig:Bound_State_Pole}
\end{figure}

\end{document}